\documentclass{article}

\usepackage[english]{babel}
\usepackage{graphicx}

\usepackage{amsmath,amsthm}
\usepackage{amsfonts}

\newcommand{\R}{\mathbb{R}}
\newcommand{\Z}{\mathbb{Z}}
\newcommand{\N}{\mathbb{N}}

\newcommand{\cylindre}{{\cal C}_r}
\renewcommand{\H}{{\cal H}}
\renewcommand{\L}{L}
\newcommand{\bPi}{{\bar{\bf\Pi}}^r}
\newcommand{\sPi}{\Pi^r}

\numberwithin{equation}{section}
\theoremstyle{plain}
\newtheorem{theorem}{Theorem}[section]
\newtheorem{corollary}[theorem]{Corollary}
\newtheorem{lemma}[theorem]{Lemma}
\newtheorem{proposition}[theorem]{Proposition}

\def\Heff{H_{\rm eff}}

\def\eff{{\rm eff}}

\def\demi{\frac{1}{2}}
\def\demi{\frac{1}{2}}

\begin{document}

\noindent 
\begin{center}
\textbf{\large Effective models for excitons in carbon
nanotubes}
\end{center}

\begin{center}
28th of April, 2006
\end{center}

\vspace{0.5cm}

\noindent 

\begin{center}
\textbf{ 
Horia D. Cornean\footnote{Dept. Math., 
    Aalborg
    University, Fredrik Bajers Vej 7G, 9220 Aalborg, Denmark; e-mail:
    cornean@math.aau.dk},
Pierre Duclos\footnote{Centre de Physique Th\'eorique UMR 6207 - Unit\'e Mixte de Recherche du CNRS et des Universit\'es Aix-Marseille I, Aix-Marseille II et de l'universit\'e du Sud Toulon-Var - Laboratoire affili\'e \`a la FRUMAM, Luminy Case 907, F-13288 Marseille Cedex 9 France; e-mail: duclos@univ-tln.fr},  
Benjamin Ricaud \footnote{Centre de Physique Th\'eorique UMR 6207 - Unit\'e Mixte de Recherche du CNRS et des Universit\'es Aix-Marseille I, Aix-Marseille II et de l'universit\'e du Sud Toulon-Var - Laboratoire affili\'e \`a la FRUMAM, Luminy Case 907, F-13288 Marseille Cedex 9 France; 
e-mail: ricaud@cpt.univ-mrs.fr}}
     
\end{center}

\vspace{0.5cm}

\begin{abstract} 
We analyse the low lying spectrum of a model of
excitons in carbon nanotubes. Consider two particles with opposite
charges and a Coulomb
self-interaction, placed on an infinitely long cylinder. If the
cylinder radius becomes small, the low lying spectrum of their
relative motion is well 
described by a one-dimensional
effective Hamiltonian which is exactly solvable.
\end{abstract}

\tableofcontents
\section{Introduction}
In order to understand the quantum mechanics of carbon nanotubes, one
has to reconsider many classical problems in which the systems are 
restricted to low dimensional configuration spaces. The effects
induced by these special shapes are significant. 
For example, optical properties and electrical conductivity in
nanotubes and nanowires are highly influenced by their geometry.
 
In a periodic structure, bands of allowed and forbidden energies are
characteristic for non-interacting electrons. When applying an
external perturbation, such as light, electrons can only absorb the
amount of energy which allows them to jump from an occupied level of
energy to a free one. In the particular case of a semi-conductor, at
low temperatures, the band of energies are either full (valence bands)
or empty (conduction bands). So the electron must absorb a fairly
large amount of energy to jump to the conduction band. 

When the self-interaction is also considered, the mathematical problem
of the optical response becomes very difficult, and there are not many
rigorous results in this direction.  Here is a sketchy description of what
physicists generally do (see for example the book of Fetter and
Walecka \cite{FW}:

\begin{enumerate}
\item Work in the grand-canonical ensemble, at zero temperature, and
  the Fermi energy $E_F$ is in 
  the middle of an energy gap;
\item Switch to an electron-hole representation, via a Bogoliubov
  unitary transformation. The new non-interacting ground state is the tensor product
  of two vacua. If before an excited state meant that an electron was
  promoted from an occupied energy state from below $E_F$ to an empty
  state above $E_F$, in the electron-hole representation it simply
  means that an electron-hole pair was created;
\item Try to diagonalize in one way or the other the true,
  self-interacting  many-body
  Hamiltonian by restricting it to a certain subspace of "physically
  relevant" excited states; this usually amounts to formulate a
  Hartree-Fock problem;
\item Try to obtain an effective one-body Hamiltonian, whose spectrum approximates in some way the original problem in a
  neighborhood of $E_F$;
\item The bound states of this effective one-body operator are called
  excitons. They describe virtual, not real particles;
\item Use the exciton energies to correct the optical response formulas
  derived in the non-interacting case. 
\end{enumerate}
Now this one-body effective Hamiltonian is a complicated object in
general. If one makes a number of further assumptions like:

\begin{enumerate}
\item There is only one conduction band above $E_F$ and only one
  valence band below $E_F$;
\item The dispersion law of these two energy bands is replaced 
  with paraboloids (effective mass approximation),
\end{enumerate}
then this one body effective Hamiltonian is nothing but the one
describing the relative motion of a positively charged particle (a
hole) and a negatively charged particle (electron), interacting
through an eventually screened Coulomb potential. Let us stress that
this procedure is generally accepted as physically sound 
in the case when the crystal is periodic
in all three dimensions.

If a special geometry is imposed (i.e. the electrons are confined on
long and thin cylinders made out of carbon atoms) then the above
procedure has to be completely reconsidered. The problem is even more
complicated, because two dimensions are on a torus and the band
structure only arises from the longitudinal variable. Even the
position of the Fermi level moves when the radius of the cylinder
varies.

It has been argued by physicists \cite{Ped1, Ped2} that 
one can still write down a Hartree-Fock type 
eigenvalue problem which describes the excited states near the
Fermi level. This operator is a two-body one, which
does not in general allow a complete reduction of its mass center. 
A mathematically sound formulation of this Hartree-Fock problem would
be of certain interest, but it is not what we do in this paper. We
rather perform the spectral analysis of an operator which has
been conjectured by physicists as being the relevant one.

The main point in investigating these low dimensional structures, is
that the interaction between electrons is enhanced and gives much
stronger exciton effects than in three dimensions. This means that
some new energy states appear deep inside the forbidden energy
band. The smaller the radius, the more important these new energies
are. That is because they allow photons with much smaller energy than 
the band gap to be absorbed into the material.

We will therefore consider two charged particles living on a cylinder
and interacting through an attractive Coulomb potential. As we have already
pointed out above, this operator models an effective Hamiltonian for
excitons in carbon nanotubes, according to~\cite{Ped1}
and~\cite{Ped2}. {\it Remember that it has nothing to do with real particles
  living on the nanotube, the exciton being just a mathematical
  artifact describing virtual particles}. We hope that our
current results could also describe excitons living in nanowires~\cite{Aki}, or two 
particles in a strong magnetic field as in~\cite{BD}. Let us mention
that our paper is an improvement and a continuation of a previous work
done in~\cite{CDP}.

\section{The mathematical model}


Our configuration space is a cylinder of radius $r$ and infinite
length, space denoted by $\cylindre=\R\times r S^1$, $S^1$ being the 
unit circle. The coordinates on the cylinder are $(x,y)\in(\R\times r
S^1)$ where $x$ is the variable along the tube axis and $y$ is the
transverse coordinate.

The two virtual particles live in the Hilbert space
$\L^2(\cylindre\times \cylindre)$. We formally consider the Hamiltonian
\begin{equation}\label{pricipessa}
\bar{H}^r=-\hbar^2\left(\frac{\Delta_{x_1}}{2m_1}+\frac{\Delta_{x_2}}{2m_2}+\frac{\Delta_{y_1}}{2m_1}+\frac{\Delta_{y_2}}{2m_2}\right)-V^r(x_1-x_2,y_1-y_2),
\end{equation}
where
\begin{equation}\label{zaculomb}
V_r(x,y):=\frac{-e_1 e_2}{\varepsilon\sqrt{x^2+4r^2\sin^2\left(\frac{y}{2r}\right)}}
\end{equation}
$(x_i,y_i)$ are the coordinates on the cylinder of the two charged
particles, $m_i$ their masses, and $e_i$ their charges which are of
opposite charge so that $V_r>0$. Here
$\varepsilon$ is the electric permittivity of the material. In the
sequel we will set $\hbar=\varepsilon=1$. The
potential $V^r$ is the three dimensional Coulomb potential simply
restricted to the cylinder. We justify the expression of $V^r$ by
Pythagora's theorem. The cylinder is embedded in $\R^3$. The distance $\rho$ from one particle to the other in $\R^3$ is:
$$\rho^2=(x_1-x_2)^2+4r^2\sin^2\left(\frac{y_1-y_2}{2r}\right)$$
where $|2r\sin\left(\frac{y_1-y_2}{2r}\right)|$ is the length of the
chord joining two points of coordinate $y_1$ and $y_2$ on the circle.

Now consider the space
\begin{align}\label{autoadj}
{\cal D}_0& =\left \{f\in
  C^{\infty}(\cylindre\times\cylindre):\quad \forall\alpha_{1,2}, 
\beta_{1,2}, \gamma_{1,2}\in \N \frac{}{}\right .\\ 
 &
 \left .\frac{}{}
|x_1^{\alpha_1}x_2^{\alpha_2}D_{x_1}^{\beta_1}D_{x_2}^{\beta_2}
D_{y_1}^{\gamma_1}D_{y_2}^{\gamma_2}f(x_1,y_1,x_2,y_2)|\le
  C_{\alpha,\beta,\gamma}\ \right \}\nonumber 
\end{align}
of "Schwartz functions along $x$" and smooth and $2\pi r$-periodic
along $y$. Clearly ${\cal D}_0$ is dense in the Sobolev space
$\H^1(\cylindre^2)$. Let us define on ${\cal D}_0$ the kinetic quadratic form
\begin{align}\label{cinetica}
t_0[\psi]=\frac{1}{2m_1}\left
  (\|\partial_{x_1}\psi\|^2+\|\partial_{y_1}\psi\|^2\right )+\frac{1}{2m_2}\left
  (\|\partial_{x_2}\psi\|^2+\|\partial_{y_2}\psi\|^2\right )
\end{align}
and the quadratic form associated with the Coulomb potential on the cylinder:
\begin{align}\label{teve}
& t_V[\psi]=\|\sqrt{V_r}\psi\|^2, \\ 
&{\cal
  D}_0\subset\left \{\psi\in \L^2(\cylindre^2),\frac{}{}\right .\nonumber \\
& \int_{\cylindre^2}
V_r(x_1-x_2,y_1-y_2)\mid\psi(x_1,x_2,y_1,y_2)\mid^2dx_1dx_2dy_1dx_2<\infty
\left .\frac{}{} \right \}.\nonumber 
\end{align}
Finally, define the form 
\begin{align}\label{formah}
t_H:=t_0-t_V\quad \mbox{\rm on the domain }{\cal D}_0.
\end{align}
The sesquilinear form induced by $t_0$ is densely defined, closable,
symmetric, non-negative, and its corresponding selfadjoint operator
$H_0$ is $-\frac{1}{2m_1}\Delta_1-\frac{1}{2m_2}\Delta_2$ with periodic boundary conditions in the $y$
variables. Its form domain is ${\cal H}^1(\cylindre^2)$, and is
essentially self-adjoint on $\mathcal{D}_0$.

\subsection{Center of mass separation in the longitudinal direction}
We introduce the total mass $M:=m_1+m_2$ and the reduced mass
$\mu:=\frac{m_1m_2}{m_1+m_2}$. Denote as usual with  
$$\left\{\begin{array}{cc}X=\frac{m_1x_1+m_2x_2}{m_1+m_2},\ x=x_1-x_2,\\ Y=y_2,\ y=y_1-y_2\end{array}\right.
$$
then,
$$\left\{\begin{array}{cc}x_1=X+\frac{m_2}{M}x,\ x_2=X-\frac{m_1}{M}x,\\ y_1=y+Y,\ y_2=Y \end{array}\right..
$$
Unfortunately, for the $y_1,\ y_2$ variables we cannot use Jacobi
coordinates because the transformation does not leave invariant the
domain of the Laplacian (the periodic boundary conditions are not preserved). That is why we use atomic coordinates $y$ and $Y$
instead. In these new coordinates, the total Hilbert space splits in a tensor
product $L^2(\R)\otimes L^2[\R\times (r S^1)^2]$. 
More precisely, if we denote by 
$$U^{-1}:L^2(\R)\otimes L^2[\R\times (r S^1)^2]\mapsto
L^2(\cylindre^2),\quad [U^{-1}f](x_1,y_1,x_2,y_2)=f(X,x,Y,y)$$
then it is quite standard to show that after this variable
change we can separate away the $X$ variable and for $f,g\in
U[\mathcal{D}_0]$ we get:
\begin{align}\label{primaschi}
t_H(U^{-1}f,U^{-1}g)&=\frac{1}{2M}\langle
  \partial_Xf,\partial_Xg\rangle +\frac{1}{2m_2}\langle
  \partial_Yf,\partial_Yg\rangle \\
&+\frac{1}{2\mu}\langle \partial_x f,\partial_xg\rangle
+\frac{1}{2\mu}\langle \partial_y f,\partial_yg\rangle \nonumber \\
&-\frac{1}{m_2}\langle \partial_y f,\partial_Yg\rangle-
\langle \sqrt{V_r(x,y)}f,\sqrt{V_r(x,y)}g\rangle.\nonumber
\end{align}
Note that the subset $U[\mathcal{D}_0]$ has the same properties as
$\mathcal{D}_0$ but in the new variables. 
Therefore we can concentrate on the reduced form 
\begin{align}\label{primaschi2}
t_h(f,g)&=\frac{1}{2m_2}\langle
  \partial_Yf,\partial_Yg\rangle \\
&+\frac{1}{2\mu}\langle \partial_x f,\partial_xg\rangle
+\frac{1}{2\mu}\langle \partial_y f,\partial_yg\rangle \nonumber \\
&-\frac{1}{m_2}\langle \partial_y f,\partial_Yg\rangle-
\langle \sqrt{V_r(x,y)}f,\sqrt{V_r(x,y)}g\rangle,\nonumber
\end{align}
densely defined on smooth enough functions in $L^2[\R\times (r
S^1)^2]$, decaying along the $x$ variable. Consider the decomposition 
$ L^2[\R\times (rS^1)^2]=\bigoplus_{k\in\Z}L^2(\R\times rS^1)$
implemented by the Fourier series 
$$ f(x,y,Y)=\sum_{k\in\Z}\frac{e^{iY\frac{k}{r}}}{\sqrt{2\pi r}}\hat f_{k,r}(x,y)
$$
where
\begin{eqnarray*}
\hat f_{k,r}(x,y)=\frac{1}{\sqrt{2\pi r}}\int_{0}^{2\pi r}f(x,y,Y)e^{-iY\frac{k}{r}}dY
\end{eqnarray*}
Then for our form $t_h$ we get:
\begin{equation}\label{sumadeforme}
t_h=:\bigoplus_{k\in \Z} t_{h_k},
\end{equation}
where $t_{h_k}$ is
\begin{align}\label{teka}
t_{h_k}(f,g)&=\frac{1}{2\mu}\langle \partial_xf,\partial_xg\rangle 
+\frac{1}{2\mu}\langle\partial_yf,\partial_yg\rangle +
\frac{k^2}{2m_2r^2}\langle f,g\rangle \\
&-\frac{ik}{ m_2r}\langle f,\partial_yg\rangle 
-\langle \sqrt{V(x,y)}f,\sqrt{V(x,y)}g\rangle.\nonumber 
\end{align}
defined on the domain ${\cal E}_k$,
$${\cal E}_k=\left\{f\in C^{\infty}(\R\times rT)\
  |x^{\alpha}D_x^{\beta}D_y^\gamma f(x,y)|\le C_{\alpha\beta\gamma},\
  \forall\alpha,\beta,\gamma \in\N\right\}.
$$
Now remember that we are only interested in the low lying spectrum of
our original operator. We will now show that for small $r$, only
$t_{h_0}$ contributes to the bottom of the spectrum. Indeed, let us
concentrate on the operator 
$$-\frac{1}{2\mu}\partial_y^2+\frac{k^2}{2m_2r^2}-i\frac{k}{m_2r^2}\partial_y$$
defined on $rS^1$ with periodic boundary conditions. Via the discrete
Fourier transformation it is unitarily equivalent to: 
\begin{align}\label{bigo2}
&\bigoplus_{p\in\Z}\left
  (\frac{1}{2\mu}\frac{p^2}{r^2}+\frac{k^2}{2m_2r^2}-\frac{kp}{m_2r^2}\right )
\end{align}
where 
\begin{align}\label{elipttic}
\frac{1}{2\mu}\frac{p^2}{r^2}+\frac{k^2}{2m_2r^2}-\frac{kp}{m_2r^2}=
\frac{1}{2r^2} (p,k)\cdot \begin{pmatrix}
\frac{1}{\mu}&-\frac{1}{m_2}\\ -\frac{1}{m_2}&\frac{1}{m_2} \end{pmatrix}\begin{pmatrix}p\\k\end{pmatrix}.
\end{align}
A simple calculation shows that both eigenvalues of the above matrix
are positive; denote with $\lambda_-$ the smaller one. Then the
operator in \eqref{bigo2} obeys:
\begin{align}\label{bigo3}
&\bigoplus_{p\in\Z}\left
  (\frac{1}{2\mu}\frac{p^2}{r^2}+\frac{k^2}{2m_2r^2}-\frac{kp}{m_2r^2}\right
  )\geq -\frac{\lambda_-}{2}\partial_y^2+\frac{\lambda_-k^2}{2r^2}.
\end{align}
Using this in \eqref{teka} we obtain the inequality 

\begin{align}\label{inegali5}
t_{h_{k}}(f,f)&\ge\min\left\{1,\mu\lambda_-\right\}\frac{1}{2\mu}[
\langle \partial_xf,\partial_x f\rangle +\langle
\partial_yf,\partial_yf)]\nonumber \\
&-\langle \sqrt{V_r(x,y)}f,\sqrt{V_r(x,y)}f\rangle +
\frac{\lambda_-k^2}{2r^2}||f||^2\nonumber \\
&=\tilde{t}_{h_{0}}(f,f)+\frac{\lambda_-k^2}{2r^2}||f||^2
\end{align}
where
$\tilde{t}_{h_{0}}$ is obviously defined by the previous line. Now one of the results
obtained in this paper will be that the spectrum of the self-adjoint
operator associated to a form like $\tilde{t}_{h_{0}}$ is bounded from
below by a numerical constant times $-(\ln(r))^2$. Hence if $k\neq 0$ and
$r$ is small enough, all $t_{h_{k}}$ will be positive and only
 $t_{h_{0}}$ will contribute to the negative part of the spectrum.


\subsection{The self-adjointness problem}
Due to \eqref{primaschi}, \eqref{primaschi2}, \eqref{sumadeforme} and
\eqref{inegali5}, it is clear that it is enough to concentrate on
$t_{h_{0}}$. If we can prove that it is bounded from below, then all
other forms with $k\neq 0$ will also have this property, and the total Hamiltonian will be
a direct sum of Friederichs' extensions. 
Because we can anyway scale the masses and charges away, and in order to simplify the
notation, let us consider the sesquilinear form:
\begin{align}\label{tetilda}
\widetilde{t_{H}}(f,g)&:=\frac{1}{2}[\langle
\partial_xf,\partial_xg\rangle
+\langle\partial_yf,\partial_yg\rangle]-\langle
\sqrt{V_r(x,y)}f,\sqrt{V_r(x,y)}g\rangle \nonumber \\
&=:\widetilde{t_0}(f,g)-\widetilde{t_V}(f,g)
\end{align}
on the domain
$${\cal E}=\left\{f\in C^{\infty}(\R\times r S^1)\
  |x^{\alpha}D_x^{\beta}D_y^\gamma f(x,y)|\le C_{\alpha\beta},\
  \alpha,\beta,\gamma \in\N\right\},
$$
now where $V_r$ is as in \eqref{zaculomb} but with $e_1e_2=-1$. 

We will now construct a self-adjoint operator out of this form.
\begin{proposition}\label{VR0compact}
The operator $V_r$ is relatively compact in the form sense with
respect to the operator $-\frac{1}{2}(\partial_x^2+\partial_y^2)$ with
form domain $\H^1(\cylindre)$. Thus the form $\widetilde{t_H}$ defines a self adjoint operator
$\widetilde{H^r}$ whose form domain is $\H^1(\cylindre)$, and $\sigma_{ess}(\widetilde{H^r})=[0,\infty)$.  
\end{proposition}

\begin{proof} We identify the cylinder $\cylindre$
with the strip $\R\times [-r\pi,r\pi]$. For every $\alpha > 0$ we define
  $\H^\alpha(\cylindre)$ to be the set of all functions which (at
  least formally) can be
  expressed as:
\begin{align}\label{tetilda222}
& f(x,y)=\frac{1}{2\pi \sqrt{r}}\sum_{m\in \Z}\int_\R e^{i
  px+imy/r}\hat{f}_m(p) \; dp,\quad \nonumber \\
& \sum_{m\in \Z}\int_\R
(1+|p|^{2\alpha}+|m|^{2\alpha})|\hat{f}_m(p)|^2 dp <\infty.
\end{align}

Let $\chi$ be the characteristic function of
the interval $[-r/2,r/2]$. Since near the boundary of the strip 
$V_r(x,y)\cdot (1-\chi(y))$ is bounded, we
only have to look at $\tilde{V}(x,y):=V(x,y)\cdot\chi(y)$. Then we can
find a constant $C$ such that everywhere in $\cylindre$ we have
$$|\tilde{V}(x,y)|\leq \frac{C}{\sqrt{x^2+y^2}}.$$
Denote with $\rho:=\sqrt{x^2+y^2}$. 
Choose a function $\chi_1\in C_0^\infty(\R)$ with support in $(-3r/2,3r/2)$,
such that $\chi\chi_1=\chi$. Then the operator of multiplication by
$\chi_1$ is 
bounded from $\H^\alpha(\cylindre)$ to $\H^\alpha(\R^2)$ and
vice versa, because it does not touch the boundary (the proof of this
fact is standard). Moreover, if $-\Delta$ is the operator
associated to $\widetilde{t_0}$, then we have 
$$\H^\alpha(\cylindre)=(-\Delta +1)^{-\alpha/2}L^2(\cylindre).$$

Note that $\mathcal{E}$ is dense in any 
$\H^\alpha(\cylindre)$. Moreover, for every $\psi\in \mathcal{E}$ we have  
\begin{equation}\label{formbound}
|\langle \psi, \tilde{V}\psi\rangle_{L^2(\cylindre)}|\leq C \langle \chi_1\psi, \frac{1}{\rho}\chi_1\psi\rangle_{L^2(\R^2)}.
\end{equation} 
We have that $\chi_1\psi\in  \mathcal{S}(\R^2)$. Then we can write 
$$\langle \chi_1\psi, (1/\rho)\chi_1\psi\rangle_{L^2(\R^2)}=\int_0^{2\pi}
\int_0^\infty \chi_1^2(\rho\sin(\theta))\overline{\psi}(\rho,\theta)\cdot \psi(\rho,\theta)d\rho d\theta$$
and after integration by parts in the radial integral we obtain

\begin{align}\label{integrparti}
&\langle \chi_1\psi, (1/\rho)\chi_1\psi\rangle_{L^2(\R^2)}\nonumber \\
&=-\int_0^{2\pi}
\int_0^\infty 
\{\partial_{\rho}[\chi_1(\rho\sin(\theta))\overline{\psi}(\rho,\theta)\cdot
\chi_1(\rho\sin(\theta))\psi(\rho,\theta)]\}\rho d\rho d\theta.
\end{align}
Then using the 
estimate $|\partial_{\rho}(\chi_1\psi)|\leq |\nabla(\chi_1\psi)|$, and with the
Cauchy-Schwarz inequality: 
\begin{align}
\langle \chi_1\psi, (1/\rho)\chi_1\psi\rangle_{L^2(\R^2)} &\leq 2
||\chi_1\psi||_{L^2(\R^2)}
||\nabla (\chi_1\psi)||_{L^2(\R^2)}\nonumber \\
&\leq {\rm const} 
||\psi||_{L^2(\cylindre)}||\psi||_{\H^1(\cylindre)}.
\end{align}
Now for an arbitrarily small $\epsilon>0$ we have
$$|\langle \psi, \tilde{V}\psi\rangle_{L^2(\cylindre)}|
\leq (C_1/\epsilon) 
||\psi||_{L^2(\cylindre)}^2+C_1\; \epsilon\; 
\widetilde{t_0}(\psi,\psi),$$
where $C_1$ is just a numerical constant. The density of $\mathcal{E}$
in $\H^1(\cylindre)$ finishes the proof of relative boundedness, and
we can define $\widetilde{H^r}$ as the Frederichs extension. 

Until now we have shown in an elementary way that 
$\sqrt{V_r}(-\Delta +1)^{-1/2}$ is bounded, but one can do much better 
than that. In \cite{Bo} it has been proved a two dimensional version of an 
inequality of Kato, which states the following:
\begin{equation}\label{bouzouina}
\langle \psi , |{\bf x}|^{-1}\psi\rangle_{L^2(\R^2)} 
\leq \frac{\Gamma(1/4)^4}{4\pi^2}
\langle \psi , \sqrt{-\Delta}\psi\rangle_{L^2(\R^2)}.
\end{equation}  
This inequality immediately implies that 
$\sqrt{V_r}:\H^{1/4}(\cylindre)\to \L^{2}(\cylindre)$ is bounded. Now let us show that the operator 
$\sqrt{V_r}\H^{1/2}(\cylindre)\to \L^{2}(\cylindre)$ is compact. We will in fact prove the 
sufficient condition that 
the operator $T:=
|{\bf x}|^{-1/2}(-\Delta +1)^{-1/2}$ defined on $L^2(\R^2)$ is compact.  

Indeed, let us denote by $\chi_n$ the characteristic function of the ball 
of radius $n>0$, centered at the origin in $\R^2$.  Then we can write:
$$T=\chi_n({\bf x})T\chi_n(-\Delta)+[(1-\chi_n)({\bf x})]T +
\chi_n({\bf x})T[(1-\chi_n)(-\Delta)].$$
First, the operator $\chi_n({\bf x})T\chi_n(-\Delta)$ is Hilbert-Schmidt (its 
integral kernel is an $L^2(\R^4)$ function), thus compact. Second, 
the sequence of operators $[(1-\chi_n)({\bf x})]T$ converges in norm to zero. 
Third, the sequence $\chi_n({\bf x})T[(1-\chi_n)(-\Delta)]$ can be expressed 
in the following way:
$$\chi_n({\bf x})T[1-\chi_n(-\Delta)]=\{\chi_n({\bf x})
|{\bf x}|^{-1/2}(-\Delta +1)^{-1/4}\}(-\Delta +1)^{-1/4}
[(1-\chi_n)(-\Delta)], $$
where the first factor is uniformly bounded in $n$, while the second one 
converges in norm to zero. Thus $T$ can be approximated in operator norm with a sequence of compact operators, hence it is compact.  

Therefore $V_r$ is a relatively compact form perturbation to $-\Delta$, 
hence the essential spectrum is stable, and the proof is over.
\end{proof}

\subsection{An effective operator for the low lying spectrum}
We will show in this section that at small $r$, the negative spectrum
of $\widetilde{H^r}$ can be determined by studying a one dimensional 
effective operator $H_{\eff}^r$. It is natural to expect that
the high transverse modes do not contribute much to the low region of
the spectrum. 

First, we separate $\widetilde{H^r}$ into different parts taking advantage of the
cylindrical geometry, that is to say, 
we represent $\widetilde{H^r}$ as a sum of orthogonal transverse modes
using the periodic boundary conditions 
along the circumference of the cylinder. 
Second, we analyze which part is relevant when the radius tends to zero.

We recall that $\widetilde{H^r}$ is formally given by 
$\widetilde{H^r}=-\frac{\Delta_x}{2}-\frac{\Delta_y}{2}-V_r$ in the
space $\L^2(\cylindre)\sim \L^2(\R)\otimes\L^2(rS^1)$. The domain
contains all $\psi\in \H^1(\cylindre)$ with the property that in
distribution sense we have 
\begin{equation}\label{domeniutilda}
\left (-\frac{\Delta_x}{2}-\frac{\Delta_y}{2}-V_r\right )\psi \in \L^2(\cylindre).
\end{equation}
This does not mean that the domain is $\H^2(\cylindre)$ because $V_r$
is too singular at the origin. 

Our problem has two degrees of freedom. We consider the orthonormal
basis of eigenvectors of $-\frac{\Delta_y}{2}$ with domain $\H^2_{\rm
  per}((-\pi r,\pi r))\sim\H^2(rS^1)$. Here, the Sobolev space $\H^2_{\rm
  per}((-\pi r,\pi r))$ denotes functions which are $2\pi r$-periodic
with first and second derivatives in the distribution sense in
$L^2$. We can write 
$$-\frac{\Delta_y}{2}=\sum_{n=-\infty}^{\infty} E_n^r \sPi_n $$
where the one dimensional projectors $\sPi_n$ are defined by
$$\sPi_n =\langle \chi_n^r,\; \cdot \rangle \chi_n^r,\quad 
\chi_n^r(y)=\frac{1}{\sqrt{2\pi r}}e^{in\frac{y}{r}}\mbox{ and
}E_n^r=\frac{n^2}{2r^2},\> n\in \Z .$$

We now introduce a family of orthogonal projectors
\begin{equation}\label{nthtransverse}
\bPi_n:={\bf 1}\otimes \sPi_n,
\end{equation}
which project from $\L^2(\cylindre)$ into what we call the $n^{th}$
{\sl transverse mode}. The operator $\widetilde{H^r}$ can be split as follows:
\begin{align}
\widetilde{H^r}=\sum_{n,m}\bPi_n \widetilde{H^r}\bPi_{m}=:
\sum_n H_n^r \otimes (\langle \chi_n^r,\; \cdot \rangle \chi_n^r)+\sum_{n\neq
  m}H_{n,m}^r \otimes (\langle \chi_m^r,\; \cdot \rangle \chi_n^r), 
\end{align}
where the sum is a direct sum, since the projectors are
orthogonal. By a natural unitary identification, we can work in a new Hilbert space:
\begin{align}\label{newhs}
\mathcal{H}=l^2[\Z; \L^2(\R)],\qquad \mathcal{H}\ni \psi
=\{\psi_n\}_{n\in\Z}, \; \psi_n\in \L^2(\R).
\end{align}
Therefore our original operator is an infinite matrix now,
$\{H_{n,m}\}_{n,m\in \Z}$ whose
elements are operators in $\L^2(\R)$. 

If $n\neq m$, the only contribution comes from $V_r$, and the
corresponding operator is a multiplication operator given by ($x\neq 0$): 
\begin{align}\label{vemn}
V_{n,m}^r(x)&:=\frac{1}{2\pi r}\int_{-\pi r}^{\pi r}V(x,y)
e^{i(m-n)\frac{y}{r}}dy,\quad x\neq 0.
\end{align}

If $n=m$, then the corresponding diagonal element is given by the operator:
\begin{align}\label{hasdia}
 H_n^r=-\frac{1}{2} \frac{d^2}{dx^2}-V^r_{\rm eff}+\frac{n^2}{2r^2},
\end{align}
where $V^r_{\rm eff}$ is deduced from $V_{n,m}^r$ when $m=n$ and is
given by 
\begin{equation}\label{potefectiv}
V_{\rm eff}^r(x)=\frac{1}{2\pi r}\int_{-\pi r}^{\pi r}V(x,y)dy.
\end{equation}
Finally, let us introduce a last notation for what will be our
effective one-dimensional comparison operator: 
\begin{equation}\label{hasefectiv}
H_{\eff}^r:=-\frac{1}{2} \frac{d^2}{dx^2}-V_{\rm eff}^r(x)
\end{equation}
and note that 
\begin{equation}\label{hasefectiv2}
H_n^r=H_{\eff}^r+\frac{n^2}{2r^2}.
\end{equation}

One can see that for $n\neq 0$, the diagonal entries of
our infinite operator valued matrix are pushed up by a term
proportional with $1/r^2$. Thus a natural candidate for a comparison
operator for the negative spectrum of $\widetilde{H^r}$ is
$H_{\eff}^r$. In the next section we will perform a careful study of
this operator.

\section{Spectral analysis of $H_{\eff}^r$}\label{section3}

We now want to study the spectrum of the operator $H_{\eff}^r$ when
$r$ becomes small. We recall that:
$$ H_{\eff}^r=-\frac{1}{2}\frac{d^2}{d x^2}-V_{\eff}^r(x)
$$
where
$$V_{\eff}^r(x)=\frac{1}{2\pi r}\int_{-\pi r}^{\pi r}\frac{1}{\sqrt{x^2+4r^2\sin^2\frac{y}{2r}}}dy
$$
with form domain $Q(H_{\eff}^r)=\H^1(\R)$. 
We are going to use perturbation theory around $r=0$, which will turn
out to be quite a singular limit. The strategy is to approximate the
form associated to the potential $V_{\eff}^r(x)$ around $r=0$ by
another quadratic form which provides a solvable approximation.

Let us define the sesquilinear form on $\mathcal{S}(\R)$ (later on we
will show that it is bounded on $\H^1(\R)$):
\begin{align}\label{thecoulombform}
& C_0(f,g):=-\int_0^{\infty} \ln(2x)\cdot
[\overline{f(x)}g(x)]'dx +\int_{-\infty}^0 \ln(-2x)\cdot
[\overline{f(x)}g(x)]'dx\nonumber \\
&=-\int_0^{\infty} \ln(x)\cdot
[\overline{f(x)}g(x)]'dx +\int_{-\infty}^0 \ln(-x)\cdot
[\overline{f(x)}g(x)]'dx +\ln(4)\overline{f(0)}g(0)\nonumber \\
&=\left \{{\rm fp} \frac {1}{|x|}+\ln(4)\;\delta \right \}(\overline{f}g).
\end{align}
The symbol $fp$ means the finite part in the sense of Hadamard, while
$\delta$ is the Dirac distribution. Note that up to an integration by parts, and for functions supported
away from zero, we have $C_0(f,g)=\langle f, \frac{1}{|x|} g\rangle$.

The main result of this subsection is contained in the following
proposition:  
\begin{proposition}\label{veff-vc}
For $r<1$ and for every $f\in\H^1(\R)$ we have the estimate
\begin{align}\label{uuu111}
\langle f,V_{\eff}^rf\rangle &=-2\ln(r)\;|f(0)|^2+C_0(f,f)
+{\cal O}\left(r^{\frac{4}{9}}\right)\cdot\|f\|^2_{\H^1(\R)}\nonumber
\\
&=-2\ln(r/2)\;|f(0)|^2+\left [{\rm fp}\frac{1}{|x|}\right ](|f|^2)
+{\cal O}\left(r^{\frac{4}{9}}\right)\cdot\|f\|^2_{\H^1(\R)}.
\end{align}
\end{proposition}

\begin{proof} The argument is a bit long, and we split it in several
  lemmas. Let us start by listing some of the properties of $V_{\eff}^r$. First
note that it scales like a "delta function", i.e. it is homogeneous of order $-1$:
\begin{equation}\label{scalingVeff}
V_{\eff}^r(x)=\frac{1}{r}V_{\eff}^1(\frac{x}{r}),
\end{equation}

The next observation is that due to the integral with respect to $y$
it is much less singular than $V_r$:
\begin{lemma}\label{logaritmicin}
The behavior of $V_{\eff}^r(x)$ is logarithmic at 0.
\end{lemma}
\begin{proof}
There exists a constant $c>0$ large enough such that for every $|y|\le\frac{\pi}{2}$ and $x\in \R$ we have
$$\frac{1}{c}(x^2+y^2)\le\frac{x^2}{4}+\sin^2(y)\le c (x^2+y^2).
$$
Thus we can integrate and obtain 
$V_{\eff}^1(x)\stackrel{x\to0}{\sim}-\ln(|x|)+{\cal O}(1)$. 
\end{proof}

\vspace{0.5cm}

We now define on $\R$ a comparison function 
$Y_r(x):=\frac{1}{\sqrt{x^2+4 r^2}}$. 
and we also denote by $Y_r$ the associated quadratic form defined on
$\H^1(\R)$. For the following, let us recall the classical Sobolev estimate in one dimension:
\begin{align}\label{Linf_H1}
||f||_\infty\leq \frac{1}{\sqrt{2}}(\|f'\|+\|f\|)=\frac{1}{\sqrt{2}}\|f\|_{\mathcal{H}_1(\R)}.
\end{align}

\begin{lemma}\label{lema22} We have the following properties:
\begin{enumerate}
\item[{\rm (i)}] $V_{\eff}^1\geq Y_1$, $V_{\eff}^1(x)-Y_1(x)={\cal
      O}(|x|^{-5}) $ for $|x|\geq 10$, and 
\begin{equation}\label{elonedif}
||V_{\eff}^1-Y_1||_{ L^1(\R)}=\ln(4);
\end{equation}

\item[{\rm (ii)}]{$\langle f,(V_{\eff}^r-Y_r) f\rangle =\ln(4)\: |f(0)|^2+{\cal O}(r^{\frac{4}{9}})\|f\|_{\H^1(\R)}^2$ $\forall r\le1$.} 
\end{enumerate}
\end{lemma}
\begin{proof}

(i). To show that $V_{\eff}^1\geq Y_1$, one uses $|\sin(\cdot)|\leq 1$. The
second estimate for $|x|\geq 10$ follows from:
\begin{align}\label{xmaimaredoi}
V_{\eff}^1(x)-Y_1(x)
&=\frac{1}{|x|}\left[\frac{1}{2\pi}\int_{-\pi
  }^{\pi}\left (1+\frac{4\sin^2\frac{y}{2}}{x^2}\right )^{-\demi}dy-
\left (1+\frac{4}{x^2}\right )^{-\demi}\right]\nonumber \\
&=\frac{1}{|x|}\left(1-\frac{1}{2\pi}\int_{-\pi
  }^{\pi}\frac{2\sin^2\frac{y}{2}}{x^2}dy+{\cal
    O}(x^{-4})-1+\frac{2}{x^2}+{\cal
    O}(x^{-4})\right),\nonumber \\
&={\cal O}(|x|^{-5}),\quad |x|\geq 10.
\end{align}
Before computing the $L^1$ norm of \eqref{elonedif}, let us notice that
none of the terms is in $L^1$. We first integrate with respect to $x$,
and then over $y$, and get:
\begin{align}
\|V_{\eff}^1-Y^1\|_{L^1(\R)}=-\frac{4}{\pi}\int_{0}^{\frac{\pi}{2}}\ln(\sin y)dy=2\ln(2),
\end{align}
thus \eqref{elonedif} is proved.

Let us now prove (ii). We have due to the scaling properties:
\begin{align}
\langle f,(V_{\eff}^r-Y_r)f\rangle &=\frac{1}{r}\int_{\R}(V_{\eff}^1-Y_1)(x/r)|f(x)|^2dx
\nonumber \\
&=\int_{\R}(V_{\eff}^1-Y_1)(x)|f(rx)|^2dx.
\end{align}
Then, we subtract the term $\|V_{\eff}^1-Y_1\|_{L^1}\cdot
|f(0)|^2=\ln(4)\cdot|f(0)|^2  $
which gives
\begin{align}
  &\langle f,(V_{\eff}^r-Y_r)f\rangle -\ln(4)\cdot
  |f(0)|^2= \nonumber \\
&=\int_{\R}(V_{\eff}^1-Y_1)(x)\left\{[f(rx)-f(0)]\overline{f(rx)}+[\overline{f(rx)}-\overline{f(0)}]f(0)\right\}dx.
\end{align}
Let $\alpha\in (0,1)$ a real number. We split the above integral in
two regions: $|x|\le r^{-\alpha}$ and $|x|\ge r^{-\alpha}$. 
We have, using \eqref{Linf_H1}:
\begin{align}
&\left | \int_{|x|\ge
    r^{-\alpha}}(V_{\eff}^1-Y_1)(x)[f(rx)-f(0)]\overline{f(rx)}dx
\right | \\
& \leq 2\|f\|_{\H^1}^2\int_{|x|\geq
  r^{-\alpha}}|(V_{\eff}^1-Y_1)(x)|dx \nonumber \\
&\leq\|f\|_{\H^1}^2\int_{|x|\ge
  r^{-\alpha}}\frac{1}{|x|^5}dx\quad\mbox{\rm if } r^{-\alpha}\geq 10
\nonumber \\
&\leq {\cal O}(r^{4\alpha})\cdot\|f\|_{\H^1}^2.
\end{align}
For the region $0\leq x \le r^{-\alpha}$ (and similarly for the other one), we can write:
\begin{align}
& \int_{0\leq x\le
  r^{-\alpha}}(V_{\eff}^1-Y_1)(x)|f(rx)-f(0)|\;|f(rx)|dx  \\
&\leq \int_{0}^{r^{-\alpha}}(V_{\eff}^1-Y_1)(x)\cdot\left |
  \int_{0}^{rx}f'(t)dt \right | \cdot |f(rx)|dx
\nonumber \\
&\le
\|V_{\eff}^1-Y_1\|_{L^1}\cdot\|f\|_{L^{\infty}}\cdot\int_{0}^{r^{1-\alpha}}|f'(t)|dt
\nonumber 
\end{align}
and the Cauchy-Schwarz inequality yields:
$$\int_{0}^{r^{1-\alpha}}|f'(t)|dt\le r^{\frac{1-\alpha}{2}}\|f\|_{\H^1}.
$$
Then we set $\alpha$ as the solution of $(1-\alpha)/2=4\alpha$ which gives $\alpha=\frac{1}{9}$.
\end{proof}

\vspace{0.5cm}

We now concentrate ourselves on $Y_r$ when $r$ is small. For the next 
two lemmas, we need to introduce the the following 
characteristic function:
\begin{equation}\label{carinte}
\chi(x)=\left\{\begin{array}{cc}1 & \mbox{\rm if } |x|\le1\\
0&\mbox{\rm otherwise}
\end{array}\right..
\end{equation}
Then we have the following lemma:
\begin{lemma}\label{veffestemar}
Consider the self-adjoint operator of multiplication by 
$\ln(|\cdot |)\chi $ defined on its natural domain in $L^2(\R)$. This 
operator is relatively bounded to $p_x:=-id/dx$, with relative bound zero.
\end{lemma}
\begin{proof} 
Indeed, $\ln(|x|)\chi(|x|)\;(p_x+i\lambda)^{-1}$, $\lambda>1$ is Hilbert-Schmidt since we have, from~\cite[XI.3]{RS3}:
\begin{align}\label{marginelog}
\|\ln(|\cdot |)\chi \;(p_x+i\lambda)^{-1}\|_{HS} &\le
{\rm const}\cdot\|\ln(|\cdot |)\chi \|_{L^2}
\|(\cdot +i\lambda)^{-1}\|_{L^2}\nonumber \\ 
&\le \frac{{\rm const}}{\sqrt{\lambda}}.
\end{align}
Note that by a similar argument as the one in \eqref{Linf_H1} 
we get the estimate: 
\begin{equation}\label{linftyldoi}
\|(p_x+i\lambda)^{-1}\|_{\L^2\mapsto \L^\infty}\leq \frac{{\rm const}}{\sqrt{\lambda}}.
\end{equation}
Then a standard argument finishes the proof. 
\end{proof}

\vspace{0.5cm}

We can now characterize the form $C_0$ introduced in \eqref{thecoulombform}:
\begin{lemma}\label{C0bounded}
The quadratic form induced by $C_0$ admits a continuous extension to
$\H^1(\R)$. Moreover, $C_0$ is infinitesimally form bounded with
respect to the form associated to $p_x^2=-d^2/dx^2$.
\end{lemma}
\begin{proof} Fix some $\epsilon \in (0,1)$. Then for every $f\in\mathcal{S}(\R)$
  we can write:
\begin{align}\label{formuleC_0}
 C_0(f,f)&=-\int_0^{\infty} \ln(2x)\cdot (d_x|f|^2)(x)dx+\int_{-\infty}^0 \ln(-2x)\cdot(d_x|f|^2)(x)dx\nonumber\\
&=\int_{-\varepsilon}^0 \ln(-2x)\cdot (d_x|f|^2)(x)dx-\int_0^{\varepsilon} \ln(2x)\cdot (d_x|f|^2)(x)dx\nonumber\\
&+\ln(2\varepsilon)\left(|f(\varepsilon)|^2+|f(-\varepsilon)|^2\right)+\int_{\R\backslash\left[-\varepsilon,\varepsilon\right]} \frac{1}{|x|}\cdot |f(x)|^2dx.
\end{align}
First we have 
\begin{eqnarray*}
\int_{\R\backslash\left[-\varepsilon,\varepsilon\right]}
\frac{1}{|x|}\cdot |f(x)|^2dx \le \frac{1}{\varepsilon}\|f\|^2.
\end{eqnarray*}
Then using \eqref{linftyldoi} we have 
\begin{equation}\label{zdrombex}
\sup_{t\in\R}|f(t)|\leq \frac{{\rm const}}{\sqrt{\lambda}}||(p_x+i\lambda)f||,
\end{equation}
which takes care of the terms containing $f(\pm \varepsilon)$. 

The remaining two integrals can be treated with the following argument:
\begin{align}\label{zdrombex2}
& \left |\int_0^{\varepsilon} \ln(2x)\cdot (d_x|f|^2)(x)dx\right |
\le 2\int_0^{\varepsilon} |\ln(2x)|\cdot |f(x)|\cdot |f'(x)|dx
\nonumber \\
&\le 2 ||\chi\; \ln(2\;|\cdot| ) f || \;\|f\|_{\H^1}
\le \frac{{\rm const}}{\sqrt{\lambda}}||(p_x+i\lambda)f||\; 
\|f\|_{\H^1},
\end{align}
where in the second inequality we used the Cauchy inequality, while in
the third inequality we used \eqref{marginelog}. 

These estimates allow us to find two constants $A$ and $B$ 
(growing when $\lambda$
grows, such that: 
$$|C_0(f,f)|\leq \frac{{\rm
    const}}{\sqrt{\lambda}}||f'||^2+A||f||\;||f'||+B||f||^2.$$
But we use the inequality $||f||\;||f'||\leq \frac{1}{\lambda
  A}||f'||^2+\lambda A||f||^2$, which finally allows us to say that
for any $0<a<1$ we can find $b>0$ such that 
\begin{equation}\label{marginirefina}
|C_0(f,f)|\leq a \;||p_x f||^2+b\;||f||^2,
\end{equation} 
 and the proof is over.
\end{proof}

\vspace{0.5cm}

The final ingredient in proving Proposition \ref{veff-vc} is contained
in the following lemma:
\begin{lemma}\label{lema33}
Recall that $Y_r(x)=(x^2+4r^2)^{-1/2}$. Then for every $r<1$, and for 
every $f\in \mathcal{S}(\R)$, we have the estimate:
$$\langle f,Y_rf\rangle =-2\ln(2r)\cdot |f(0)|^2+C_0(f,f)+{\cal
  O}\left(r^{\frac{4}{9}} \right)\cdot \|f\|_{\H^1(\R)}^2.
$$
\end{lemma}
\begin{proof}
Integrating by parts we obtain:
\begin{align}
\langle f,Y_r f\rangle &=
-2\ln(2r)|f(0)|^2\nonumber \\
& -\int_{0}^{\infty}\ln(x+\sqrt{x^2+4 r^2})
\cdot[f'(x)\overline{f(x)}+f(x)\overline{f'(x)}]dx\nonumber \\
&+\int_{-\infty}^{0}\ln(-x+\sqrt{x^2+4 r^2})\cdot[f'(x)\overline{f(x)}+f(x)\overline{f'(x)}]dx
\end{align}
and:
\begin{align}
&\langle f,Y_r f\rangle -C_0(f,f)+2\ln(2r)\cdot |f(0)|^2=\\ 
&-\int_{0}^{\infty}[\ln(x+\sqrt{x^2+4 r^2})-\ln(2x)]
\; (d_x|f(x)|^2) dx \nonumber \\
&+\int_{-\infty}^0[\ln(-x+\sqrt{x^2+4
  r^2})-\ln(-2x)]\; (d_x |f(x)|^2)dx.\nonumber
\end{align}
The idea is to show that the last two integrals are small when $r$ is
small. We only consider the first integral, since the argument is
completely analogous for the second one. 

Fix some $0<\alpha < 1$ (its optimal value will be chosen later), and
assume that $r$ is small enough such that $r^{1-\alpha}\leq 1/10$. 
We split the domain of integration into two regions: one in which 
$x>r^{\alpha}$,  and the other one where $x\leq r^\alpha$. For the
first region we have: 
\begin{align}
\ln(x+\sqrt{x^2+4
  r^2})-\ln(2x)&=\ln\frac{1+\sqrt{1+4\frac{r^2}{x^2}}}{2}={\cal
  O}(r^2/x^2).
\end{align}
Then by integration, and using \eqref{Linf_H1} together with the
Cauchy inequality, we get:
\begin{align}\label{uuun}
& \left | \int_{r^{\alpha}}^{\infty}[\ln(x+\sqrt{x^2+4 r^2})-\ln(2x)]
  f'(x)\overline{f(x)}dx\right | \nonumber \\ 
&\le 
\left | \int_{r^{\alpha}}^{\infty}[\ln(x+\sqrt{x^2+4 r^2})-\ln(2x)]
  f'(x)dx\right | \|f\|_{\H^1} \nonumber \\
&\le \left(\int_{r^{\alpha}}^{\infty}[\ln(x+\sqrt{x^2+4
  r^2})-\ln(2x)]^2dx\right)^{\demi}\|f\|_{\H^1}^2 \nonumber \\
&\le{\cal O}(r^{2-\frac{3}{2}\alpha})\|f\|_{\H^1}^2.
\end{align}
For the region where $x\le r^{\alpha}$, we use the monotonicity of the 
logarithm and write:
\begin{align}
|\ln(x+\sqrt{x^2+4 r^2})-\ln(2x)|&\le |\ln(x+\sqrt{x^2+4
  r^2})|+|\ln(2x)|\nonumber \\
&\le |\ln (2r)|+|\ln (2x)|.
\end{align}
Then we can write
\begin{align}\label{unuda}
&\left |\int_0^{r^{\alpha}}[\ln(x+\sqrt{x^2+4 r^2})-\ln(2x)] 
f'(x)\overline{f(x)}dx\right |\nonumber \\
&\le {\rm const}\cdot\|f\|_{\H^1}^2\left(\int_0^{r^{\alpha}}
[\ln(2x)+\ln (2r)]^2dx\right)^{\demi}= 
{\cal O}(r^{\frac{\alpha}{2}}|\ln r|)\cdot \|f\|_{\H^1}^2.
\end{align}
Comparing \eqref{uuun} and \eqref{unuda}, 
we see that we can take $\alpha$ arbitrarily close to $1$. In
particular, we can find some $\alpha$ such that $2-3\alpha/2> 4/9$ and
$\alpha/2>4/9$ and we are done.
\end{proof}
We can now conclude the proof of Proposition \ref{veff-vc} by putting
together the estimates from Lemma \ref{lema22} and Lemma
\ref{lema33}.

\end{proof}

\vspace{0.5cm}

\noindent {\bf Remark}. One can improve the exponent $4/9$ in the
error estimate \eqref{uuu111}, and obtain $1/2$ instead. One observes
that the Fourier transform $\widehat{V_{{\rm eff}}^1}(p)$ can be exactly
computed in terms of modified Bessel functions, and then one expands
it near $p=0$ identifying the Fourier transforms of ${\rm
  fp}\frac{1}{|x|}$ and $\delta$. Then the error's Hilbert-Schmidt
norm is estimated and shown to be of order $r^{1/2}$. A related problem treated with this method can be found in \cite{BD2}.  


\subsection{A solvable comparison operator $H_C$}

Remember that we are interested in the negative spectrum of $H_{{\rm
    eff}}^r$, operator given in \eqref{hasefectiv}. Lemmas \ref{logaritmicin}
    and \ref{veffestemar} tell us that its operator domain is
    $\H^2(\R)$, while the form domain is $\H^1(\R)$.  Proposition
    \ref{veff-vc}, see \eqref{uuu111},  indicates that a good approximation for $H_{{\rm
    eff}}^r$ at small $r$ would be the operator formally defined as 
\begin{equation}\label{hascdef}
H_C:=\frac{1}{2}p_x^2+2\ln(r/2)\;\delta -\frac{1}{|x|}.
\end{equation}
Of course, as it is written above $H_C$ makes no sense. The correct
definition of $H_C$ can be found in the Appendix A of \cite{BD2} in a
more general setting. For the comfort of the reader we give below a
version of this definition adapted to our simpler situation. $H_C$ has to be
understood in the following way: consider the sesquilinear form on
$\mathcal{S}(\R)$ given by  
\begin{align}\label{formaluihc}
t_C(f,g):=\frac{1}{2}\langle f',g'\rangle +2\ln(r)\; 
\overline{f}(0)g(0)-C_0(f,g). 
\end{align}
A standard consequence of \eqref{zdrombex} and \eqref{marginirefina}
is that the quadratic form associated with $t_C$ is closable, bounded
from below, and the domain of its closure is $\H^1(\R)$. Then $H_C$ is
the self-adjoint operator generated by $t_C$, and its operator domain
$D_C$ 
is characterized by:
   \begin{align}\label{domeniulluihc}
D_C:=\{\psi\in\H^1(\R):\; |t_C(\phi,\psi)|\leq {\rm const}\;
||\phi||,\quad \forall \phi\in \H^1(\R)\}.
\end{align}
Moreover, if $\psi\in D_C$, then we have the equality:
 \begin{align}\label{domeniulluihc2}
t_C(\phi,\psi)=\langle \phi, H_C\psi \rangle ,\quad \forall \phi\in \H^1(\R)\}.
\end{align}
Another representation for $\psi\in D_C$ is that there exists
$f_\psi=H_C\psi \in L^2(\R)$ such that the distribution $\psi''$ is a regular
distribution on $\R\setminus\{0\}$ and is given by:
 \begin{align}\label{domeniulluihc3}
\psi''(x)=-2\frac{1}{|x|}\psi(x) -2f_\psi(x).
\end{align}
One important consequence is that $\psi'\in \H^1(\R\setminus\{0\})$,
and $\psi'$ is continuous on $ \R\setminus\{0\}$.

Let us now introduce the parity operators $P_+$ and $P_-$
\begin{align}\label{paritate1}
P_{\pm}:\H^1(\R)\to\H^1(\R),\quad 
f(x)\mapsto (P_{\pm}f)(x)=\frac{f(x)\pm f(-x)}{2}.
\end{align}
We have that $P_++P_-=1$. We will call ${\rm Ran} P_+$ the {\it even
  sector} and ${\rm Ran} P_-$ the {\it odd sector}. The following
  lemma is an easy application of definitions, and we give it without proof:  
\begin{lemma}\label{parity22}
We have that $t_C(P_{\pm} f,P_{\mp}g)=0$ for all
$f,g\in\H^1(\R)$. Then the domain $D_C$ of $H_C$ is left invariant by 
$P_{\pm}$; moreover, $H_C$ commutes with $P_{\pm}$. 
\end{lemma}
\qed

A standard consequence of the elliptic regularity (see also
\eqref{domeniulluihc3}) is the following lemma, given again without proof:
\begin{lemma}
The eigenvectors of $H_C$ belong to $C^{\infty}(\R\backslash\{0\})$.
\end{lemma}

\qed 
\vspace{0.5cm}

A less obvious result is the following characterization of $D_C$: 
\begin{lemma}\label{boundarycond}
Every $\psi$ in $D_C$ obeys the following boundary condition at $0$:
\begin{equation}\label{boundcon2}
\lim_{\varepsilon\to 0} \left[\frac{\psi'(-\varepsilon)-
\psi'(\varepsilon)}{2}+2\ln(r)\psi(0)-2\ln(2\varepsilon)\psi(0)\right]=0.
\end{equation}
\end{lemma}
\begin{proof} Now remember that $\psi'$ is continuous outside the
  origin if $\psi\in D_C$, so $\psi'(\pm\varepsilon)$ makes sense.
  Moreover, for every $\phi\in \H^1(\R)$ we have:
\begin{align}\label{eqHC}
\demi\int_{\R}\overline{\phi'(x)}
\psi'(x)dx+2\ln(r)\overline{\phi(0)}\psi(0)-C_0(\phi,\psi)=\langle
\phi,H_C\psi\rangle .
\end{align}
We can write for $x>0$ (see \eqref{domeniulluihc3}):
$$\psi'(x)=\psi'(1)+\int_{1}^x\psi''(y)dy=
\psi'(1)-2\int_{1}^x\left (\frac{\psi(y)}{y}+f_\psi(y)\right )dy
$$
and then for $x$ close to $0$ we have:
\begin{equation}\label{logaloga}
|\psi'(x)|\le {\rm const}+2|\ln x|\;||\psi||_\infty.
\end{equation}
The same estimate is true for negative $x$ near $0$, and together with
the estimate \eqref{Linf_H1} it follows that $\psi'$ diverges at $0$ not 
faster than a logarithm. 

Now we can integrate by parts outside the origin and write:
\begin{align}\label{uberkul1}
& \int_{\R}\overline{\phi'(x)}\psi'(x)dx\\
&=\overline{\phi(-\varepsilon)}\psi'(-\varepsilon)-\int_{-\infty}^{-\varepsilon}\overline{\phi(x)}\psi''(x)dx-\overline{\phi(\varepsilon)}\psi'(\varepsilon)-\int_{\varepsilon}^{\infty}\overline{\phi(x)}\psi''(x)dx\nonumber
\\
&+\int_{-\varepsilon}^{\varepsilon}\overline{\phi'(x)}\psi'(x)dx\nonumber
\end{align}
where the last integral will converge to zero with $\varepsilon$. 

After a similar integration by parts we obtain:
\begin{align}\label{uberkul2}
& C_0(\phi,\psi)=\int_{-\varepsilon}^0 \ln(-2x)\cdot (d_x(\overline{\phi}\psi))(x)dx-\int_0^{\varepsilon} \ln(2x)\cdot (d_x(\overline{\phi}\psi))(x)dx\nonumber\\
&+\ln(2\varepsilon)\left(\overline{\phi(\varepsilon)}\psi(\varepsilon)+\overline{\phi(-\varepsilon)}\psi(-\varepsilon)\right)+\int_{\R\backslash\left[-\varepsilon,\varepsilon\right]} \frac{1}{|x|}\cdot \overline{\phi(x)}\psi(x)dx.
\end{align}
Following the reasoning in \eqref{zdrombex2}, one can prove that :
$$\int_0^{\varepsilon} \ln(2x)\cdot [d_x(\overline{\phi}\psi)](x)dx\stackrel{\varepsilon\to0}{=}{\cal O}(\varepsilon^{\demi}|\ln\varepsilon|)\;||\phi||_{\H^1}
||\psi||_{\H^1}$$
and thus:
\begin{align}\label{uberkul3}
C_0(\phi,\psi)\stackrel{\varepsilon\to0}{=}
\ln(2\varepsilon)\left(\overline{\phi(\varepsilon)}\psi(\varepsilon)+\overline{\phi(-\varepsilon)}\psi(-\varepsilon)\right)+\int_{\R\backslash\left[-\varepsilon,\varepsilon\right]} \frac{1}{|x|}\cdot \overline{\phi(x)}\psi(x)dx.
\end{align}
Putting \eqref{uberkul3} and \eqref{uberkul1} in \eqref{eqHC}, and
using \eqref{domeniulluihc3}, we eventually get:
\begin{align}
\lim_{\varepsilon\to 0}
& \left[\frac{1}{2}\left
    (\overline{\phi(-\varepsilon)}\psi'(-\varepsilon)-\overline{\phi(\varepsilon)}\psi'(\varepsilon)\right )+2\ln(r)
  \overline{\phi(0)}\psi(0) \right .\nonumber\\ 
&\left .\frac{}{}-\ln(2\varepsilon)\left(\overline{\phi(\varepsilon)}\psi(\varepsilon)+\overline{\phi(-\varepsilon)}\psi(-\varepsilon)\right)\right]=0.
\end{align}
The last ingredient is the embedding $\H^1(\R)\subset
C^{1/2-\delta}(\R)$, and the estimate \eqref{logaloga} which allows us
to simplify the limit: 
\begin{equation}
\lim_{\varepsilon\to0}\overline{\phi(0)}\left[\frac{\psi'(-\varepsilon)-\psi'(\varepsilon)}{2}+2\ln(r)\psi(0)-2\ln(2\varepsilon)\psi(0)\right]=0
\end{equation}
for all $\phi\in\H^1(\R)$. The lemma is proved. 
\end{proof}

\subsection{The eigenvalues and eigenvectors of $H_C$}\label{eigen}

In this subsection we give analytic expressions for eigenvalues and
eigenvectors of $H_C$ corresponding to the negative, discrete spectrum; much of the information we need about special
functions is borrowed from \cite{L}. We want to have the same formal
expression for our eigenvalue problem outside $z=0$ as in that paper, namely
$$\frac{d^2}{dz^2}\psi-\frac{1}{4}\psi+\frac{\alpha}{|z|}\psi=0
$$
where $\psi$ will be an eigenfunction with an associated eigenvalue
$E=-\frac{1}{2\alpha^2}$. Let us now do this in a rigorous manner. 

We want to implement the change of variables $x=\demi\alpha z$, $
\alpha>0$, which amounts to defining a unitary operator on $L^2(\R)$:
\begin{equation}\label{Utransf}
(U_W\psi)(z)=\sqrt{\frac{\alpha}{2}}\ \psi(\alpha z/2),\quad (U_W^{-1}\psi)(z)=\sqrt{\frac{2}{\alpha}}\ \psi(2 z/\alpha).
\end{equation}
Now assume $\phi$ is a normalized eigenvector for $H_C$ satisfying 
\begin{equation}\label{ecdevecp}
H_C\phi =E\phi,\quad E<0.
\end{equation}
Instead of solving the above equation, we will reformulate it in terms
of $\psi=U_W\phi$, and $\phi=U_W^{-1}\psi$. To do that, we need to fulfill two
conditions. The first one is:
\begin{align}\label{eqqquuu}
[U_WHU_W^{-1}\psi](z)&=\frac{2}{\alpha^2}[-\psi''(z)-\frac{\alpha}{|z|}\psi(z)]=E\psi(z),\quad
z\neq 0, \quad {\rm or}\nonumber \\
& \psi''(z)-\frac{1}{4}\psi(z)+\frac{\alpha}{|z|}\psi(z)=0,\quad
z\neq 0,\; E=-\frac{1}{2\alpha^2}.
\end{align}
The second thing is to see what condition at $z=0$ should $\psi$ obey
in order to be sure that $U_W^{-1}\psi$ is in the domain of $H_C$. If
we replace $\psi$ in \eqref{boundcon2} by $\phi=U_W^{-1}\psi$, then we get the
modified condition:

\begin{align}\label{boundcon22}
\lim_{\varepsilon\to0}\left[\frac{2}{\alpha}\frac{\psi'(-\frac{2}{\alpha}\varepsilon)-\psi'(\frac{2}{\alpha}\varepsilon)}{2}+2\ln(r)\psi(0)-
  2\ln(2\varepsilon)\psi(0)\right]=0
\end{align}
or
\begin{equation}\label{conditionalfa}
\lim_{\varepsilon\to0}\left[\frac{\psi'(-\varepsilon)-\psi'(\varepsilon)}{2}+\alpha(\ln r-\ln(\alpha\varepsilon))\psi(0)\right]=0.
\end{equation}
Therefore we reduced the problem of finding the eigenfunctions and
eigenvalues of $H_C$ to solving the ordinary differential equations in
\eqref{eqqquuu}, with the boundary condition given in
\eqref{conditionalfa}. We will see that $L^2$ solutions can be
constructed only if $\alpha$ obeys some conditions. 

A priori, $\alpha$ can be any positive real number. First assume that
$\alpha$ is not a positive integer. Then if we solve \eqref{eqqquuu}
for $z>0$, we see that the only square integrable solution at
$+\infty$ is the one given by a Whittaker function:
\begin{equation}\label{whittake}
W_{\alpha,\frac{1}{2}}(z)=ze^{-\demi z}U(1-\alpha,2,z),
\end{equation}
where $U$ is the confluent hypergeometric function or Kummer function,
see~\cite{AS}. If $\alpha$ is a positive integer, the solution is
obtained as the limit of  $W_{\alpha,\frac{1}{2}}(z)$ when $\alpha$ tends to a positive integer $N$ and get:
\begin{equation}\label{pozit22}
\lim_{\alpha\to N}W_{\alpha,\frac{1}{2}}(z)=e^{-\demi z}z\frac{1}{N}L_{N-1}^1(z)
\end{equation}
where $L_{N-1}^1$ is an associated Laguerre polynomial. 

We denote with $\Gamma(z)$ and $\psi(z)=\Gamma'(z)/\Gamma(z)$ the
usual gamma and digamma functions. We have the following first result:
\begin{proposition}\label{nedege1}
{\rm (i)}. All negative eigenvalues of $H_C$ are non-degenerate. The
eigenfunctions of $H_C$ are also eigenfunctions of $P_\pm$.

{\rm (ii)}.  There
exists an infinite number of odd eigenfunctions $\phi_{odd,k}$, $k\in\Z_+$, corresponding to every
$\alpha\in \{1,2,\dots\}$. 

{\rm (iii)}. There also exists an infinite number of
even eigenfunctions $\phi_{even,k}$, $k\in\Z_+$, each corresponding to a certain $\alpha_k\in (k-1,k)$
for every $k\in \Z_+$.   
\end{proposition}
\begin{proof}  (i). Choose any eigenfunction $\phi$ of $H_C$ corresponding
  to $E<0$. Make the change $\psi=U_W^{-1}\phi$, and then look at the
  associated  differential equation:
\begin{align}\label{eqqq2}
& \psi''(z)-\frac{1}{4}\psi(z)+\frac{\alpha}{|z|}\psi(z)=0,\quad z\neq
0,\; E=-\frac{1}{2\alpha^2}, \\
& \label{yre2}\lim_{\varepsilon\to0}\left[\frac{\psi'(-\varepsilon)-\psi'(\varepsilon)}{2}+\alpha(\ln r-\ln(\alpha\varepsilon))\psi(0)\right]=0.
\end{align}
First assume $\alpha =N\in\Z_+$. The theory of ordinary
differential equations insure the existence of two constants $C_1$ and
$C_2$ such that 
\begin{align}\label{rere3}
\psi(z)&=C_1 e^{-\demi z}z\frac{1}{N}L_{N-1}^1(z),\quad z>0,\nonumber
\\
\psi(z)&=C_2 e^{\demi z}z\frac{1}{N}L_{N-1}^1(-z),\quad z<0.
\end{align}   
By inspection (and by continuity) we get that $\psi(0)=0$. If we put
$\psi$ in \eqref{yre2}, and using the explicit form of the Laguerre
polynomials, we get that the boundary condition is fulfilled only if
$C_1=C_2$ which amounts to $\psi(z)=-\psi(-z)$, i.e. there is one and only one
solution which is also odd. 

Now assume $\alpha\not\in \Z_+$. The theory of ordinary
differential equations insure the existence of two constants $C_3$ and
$C_4$ such that 
\begin{align}\label{rere4}
\psi(z)&=C_3 W_{\alpha,\frac{1}{2}}(z),\quad z>0,\nonumber
\\
\psi(z)&=C_4 W_{\alpha,\frac{1}{2}}(-z),\quad z<0.
\end{align}   
By inspection we see that
$W_{\alpha,\frac{1}{2}}(0)=\frac{1}{\Gamma(1-\alpha)}\neq 0$, hence by
continuity at zero we must have $C_3=C_4$, hence $\psi(z)=\psi(-z)$
and the eigenfunction must be even. 

(ii). The proof is already contained in (i), since the boundary
condition is trivially fulfilled for odd functions. There is exactly
one eigenfunction, an odd one, corresponding to every $\alpha\in\Z_+$.

(iii). We saw in (i) that if there are eigenfunctions corresponding to
$\alpha\not\in\Z_+$, then they must be even. In order to get all possible $\alpha$'s which are compatible with the boundary condition
\eqref{yre2}, we compute (note that $\psi'$ is odd):
\begin{align}\label{boco44}
\lim_{\varepsilon\to0}\left[-W_{\alpha,\frac{1}{2}}'(\varepsilon)+\alpha(\ln
  r-\ln(\alpha\varepsilon))\frac{1}{\Gamma(1-\alpha)}\right]=0
\end{align}
and using the explicit expression of these special functions we obtain
the condition on $\alpha$:
\begin{equation}\label{conditioneven}
f(\alpha,r):=\psi(1-\alpha)+2\gamma+\frac{1}{2\alpha}-\ln\alpha+\ln r=0.
\end{equation}
where $\psi$ here means the digamma function and $\gamma$ is Euler's
constant. Since the digamma function is strictly increasing from
$-\infty$ to $+\infty$ on each interval of the form $(-m,-m+1)$,
$m\in\Z_+$, one
can easily see that $f(\cdot,r)$ is strictly decreasing from $+\infty$
to $-\infty$ when $\alpha$ varies in an interval of the form $(k-1,k)$
for every $k\in\Z_+$. Therefore we have a unique solution $\alpha_k\in
(k-1,k)$ of the equation $f(\alpha_k,r)$ for every $k\in\Z_+$. The
proposition is proved.

\end{proof}

\vspace{0.5cm}

The previous proposition stated that only the eigenvalues from the even
sector can vary with $r$. Let us now further investigate this
dependence.  

\begin{corollary} {\rm (i)}. The excited states with even parity tend
  to those with odd parity when $r$ is small. More precisely, for $k\geq 2$, we have that $\lim_{r\to
    0}\alpha_k=k-1$; 

 {\rm (ii)}. For $k=1$, we have the following asymptotic behavior of
 the ground state:
\begin{equation}\label{asym100}
\alpha_1(r)=-\frac{1}{2\ln(r)}\{1+{ o}_r(1)\},\quad E_1(r)=-2[\ln(r)]^{2}\{1+{ o}_r(1)\}.
\end{equation}
\end{corollary}

\begin{proof} (i). The limit follows easily from the properties of the
  digamma function.

(ii). We apply the implicit function theorem. Define the function 
$$F(\alpha,y):=\frac{2\alpha}{1+2\alpha [2\gamma
  +\psi(1-\alpha)-\ln(\alpha)]}-y,$$
for $(\alpha,y)$ in a small disk around the origin in $\R^2$. This
function is $C^1$ near $(0,0)$, $(\partial_\alpha F)(0,0)=2$, and
$F(0,0)=0$. Then for every $y>0$ small enough there exists
$\alpha(y)>0$ 
such that $F(\alpha(y),y)=0$. Now put $y=-1/\ln(r)$ and we are done
because \eqref{conditioneven} is also satisfied with this $\alpha$.
\end{proof}

\vspace{0.5cm}

\subsection{Approximation of $H_{\eff}^r$ by $H_C$}
We will now show that the negative spectrum of $H_{\eff}^r$ converges in
a certain sense to the one of $H_C$. This is made precise in the next
proposition, but before we need a definition. For a given subset $S$
of $\R$, and for any $\epsilon>0$ we define 
\begin{equation}\label{esepsilon}
S_\epsilon:= \bigcup _{x\in S}B_\epsilon(x).
\end{equation}
If $S$ is a discrete, finite set, then $S_\epsilon$ is a finite union of
intervals of length $2\epsilon$, centered at the points of $S$. 
\begin{proposition}\label{propohasc} 
The following three statements hold true:

{\rm (i)}. Fix $a<0$, and denote by $A:=\sigma(H_C)\cap
  (-\infty,a]$ and $B:=\sigma(H_{\eff}^r)\cap
  (-\infty,a]$. Then for every $\epsilon>0$, there exists
  $r_\epsilon>0$ such that for every $r<r_\epsilon$ we have
\begin{equation}\label{dist333}
A\subset B_\epsilon,\quad B\subset A_\epsilon.
\end{equation}

{\rm (ii)}. The ground-state of $H_{\eff}^r$ is non-degenerate, has even parity, and
diverges to $-\infty$ when $r\to 0$. Moreover:

\begin{equation}\label{dist334}
\lim_{r\to 0}|\inf\sigma(H_C)-\inf\sigma(H_{\eff}^r)|=0.
\end{equation}

{\rm (iii)}. Fix a compact interval $[a,b]\subset (-\infty,0)$ and
suppose that $H_C$ has exactly one eigenvalue of a given parity $E_C$
in $[a,b]$, for all $r<r_0$. Then if $r$ is small
enough, $H_{\eff}^r$ has exactly one
eigenvalue of the same parity $E_{\eff}$ in this interval and 
$$\lim_{r\to 0}|E_{\eff}-E_C|=0.$$

\end{proposition}

\begin{proof}Let us introduce the resolvents $R_{\eff}(z)=(H_{\eff}^r-z)^{-1}$ for
all $z\in\rho(H_{\eff}^r)$ and $R_{C}(z)=(H_{C}-z)^{-1}$ for all
$z\in\rho(H_C)$. The first ingredient in the proof is contained by the
following lemma:

\begin{lemma}\label{hasefhasr}
There exists a constant $K>1$ sufficiently large, and $r_0$
small enough, such that for every $r<r_0$ we
have that the form defined on $L^2(\R)\times L^2(\R)$ (see also 
\eqref{formaluihc})
\begin{align}\label{vetildar}
V_C(f,g)&:=t_C\left ([p_x^2/2+\lambda_r ]^{-1/2}f ,[p_x^2/2+\lambda_r
]^{-1/2}g\right )\nonumber \\
& +\lambda_r\langle f,[p_x^2/2+\lambda_r ]^{-1}g\rangle
-\langle f,g\rangle\nonumber \\
&"="[p_x^2/2+\lambda_r ]^{-1/2}\{H_C+\lambda_r\}[p_x^2/2+\lambda_r
]^{-1/2}-{\rm Id},\quad \lambda_r:=K \ln^2(r).
\end{align}
generates a bounded operator on $L^2(\R)$ denoted in the same way. Moreover, $\sup_{0<r<r_0}||V_C||\leq 1/2$. And we have 
\begin{equation}\label{vetildar2}
\{H_C+\lambda_r\}^{-1}=[p_x^2/2+\lambda_r ]^{-1/2}\{{\rm Id}+V_C\}^{-1}[p_x^2/2+\lambda_r]^{-1/2}.
\end{equation}
\end{lemma}

\begin{proof} The key estimate is contained in 
\begin{equation}\label{bokaboka33}
||(p_x^2/2+\lambda)^{-1/2}||_{L^2\to L^\infty}\leq {\rm const}\;\frac{1}{\lambda^{1/4}},
\end{equation}
obtained with an argument as in \eqref{linftyldoi}. Then if we have
$|\ln(r)|/\sqrt{\lambda_r}$ small enough, then the "delta function" part
of $t_C$ will be small uniformly in $r<r_0$. Using also \eqref{marginirefina}, and the
definition \eqref{formaluihc}, then one can show that \eqref{vetildar}
is a bounded sesquilinear form on $L^2(\R)$, with a norm which can be
made arbitrarily small if $K$ is chosen large enough. Now the equality
\eqref{vetildar2} is easy, and note that this is also compatible with \eqref{asym100} . 
\end{proof}

Introduce the notation:
\begin{align}\label{vetildar34}
\tilde{V}_{\eff}:=(p_x^2/2+\lambda_r)^{-1/2}V_{\eff}^r(p_x^2/2+\lambda_r)^{-1/2}.
\end{align}
The second ingredient in the proof of the above proposition is the
following estimate, which is an easy consequence of Proposition
\ref{veff-vc}:
\begin{align}\label{vetildar33}
||V_C-\tilde{V}_{\eff}||=\mathcal{O}(r^{4/9}),\quad r<r_0.
\end{align}
We also have that $||\tilde{V}_{\eff}||\leq 2/3$ if $r_0$ is
small enough, uniformly in $r<r_0$, and then 
\begin{equation}\label{vetildar208}
\{H_{\eff}^r+\lambda_r\}^{-1}=[p_x^2/2+\lambda_r ]^{-1/2}\{{\rm Id}+\tilde{V}_{\eff}\}^{-1}[p_x^2/2+\lambda_r]^{-1/2}.
\end{equation}
It is clear that a similar identity would hold for any other
$\lambda\geq \lambda_r$, and this already tells us that the spectrum
of $H_{\eff}^r$ is contained in an interval of the type
$(-\lambda_r,\infty)$, thus justifying the discussion after
\eqref{inegali5}. 

From \eqref{vetildar208}, \eqref{vetildar33} and
\eqref{vetildar2}, we get the crucial estimate:
\begin{equation}\label{vetildar209}
||R_{\eff}(-\lambda_r)-R_C(-\lambda_r)||\leq {\rm
  const}\;\frac{r^{4/9}}{\lambda_r},\quad r<r_0.
\end{equation}
This estimate allows us to prove (i). Introduce the notation 
\begin{equation}\label{vetildar210}
d_C(z):={\rm dist}(z,\sigma(H_C)).
\end{equation}
Choose $z\in\rho(H_C)$ (thus $d_C(z)>0$). From the identity:
\begin{equation}\label{vetildar2311}
(H_C-z)R_C(-\lambda_r)={\rm Id}-(z+\lambda_r)R_C(-\lambda_r)
\end{equation}
we get that the right hand side is invertible and:
\begin{align}\label{vetildar211}
\left \{{\rm Id}-(z+\lambda)R_C(-\lambda_r)\right
\}^{-1}&=(H_C+\lambda_r)R_C(z)={\rm Id}+(z+\lambda_r)R_C(z).
\end{align}
The first equality implies that 
\begin{align}\label{vetildar212}
R_C(z)=R_C(-\lambda_r)\left \{{\rm Id}-(z+\lambda_r)R_C(-\lambda_r)\right
\}^{-1},
\end{align}
while the second one gives the norm estimate:
\begin{align}\label{vetildar213}
\left \Vert \left \{{\rm Id}-(z+\lambda_r)R_C(-\lambda_r)\right
\}^{-1}\right \Vert \leq [1+(|z|+\lambda_r)/d_C(z)].
\end{align}
Note the important fact that \eqref{vetildar212} is just another form
of the resolvent identity, valid for any self-adjoint operator. If we
could replace $R_C(-\lambda_r)$ by $R_{\eff}(-\lambda_r)$, then the
right hand side would immediately imply that $z\in \rho(H_{\eff})$. 

We can restrict ourselves to those $z$'s which obey $|z|\leq
\lambda_r$. Then using \eqref{vetildar209} and \eqref{vetildar213}, we
get that for $r<r_0$ and $d_C(z)\geq r^{1/3}$, the operator 
$${\rm Id}-(z+\lambda_r)R_{\eff}(-\lambda_r)$$ is invertible and we get
the estimate:
\begin{align}\label{vetildar214}
\left \Vert \left \{{\rm Id}-(z+\lambda_r)R_{\eff}(-\lambda_r)\right
\}^{-1}\right \Vert \leq \frac{{\rm const}}{d_C(z)}\;\lambda_r\; 
[1-{\rm const}\;\lambda_r \; r^{4/9}/d_C(z)]^{-1}.
\end{align}
Therefore we have proved that for every $z$ which obeys $|z|\leq
\lambda_r$ and $d_C(z)>r^{1/3}$, the operator 
\begin{align}\label{vetildar215}
R_{\eff}(-\lambda_r)\left \{{\rm Id}-(z+\lambda_r)R_{\eff}(-\lambda_r)\right
\}^{-1}
\end{align}
exists and defines $R_{\eff}(z)$. It means that the spectrum of
$H_{\eff}^r$ is "close" to that of $H_C$, and the distance between them
is going to zero at least like $r^{1/3}$. 

Let us now prove (ii). We know that the ground-state of $H_C$ diverges
like $-\ln^2(r)$ for small $r$, and it is isolated from the rest of
the spectrum. Choose a circular contour $\Gamma$ of radius $1$ around this
ground-state. It means that $d_C(z)=1$ for $z\in\Gamma$. 
Then \eqref{vetildar215}, \eqref{vetildar214} and \eqref{vetildar209}
imply the estimate 
\begin{align}\label{vetildar216}
\sup_{z\in \Gamma}||R_{\eff}(z)-R_C(z)||\leq {\rm const}\cdot 
r^{4/9},\quad r<r_0.
\end{align}
Now we can employ the regular perturbation theory, see \cite{K}, 
by using Riesz
projections defined as complex integrals of the resolvents 
on contours like $\Gamma$. Then the estimate \eqref{dist334} is
straightforward. 

Finally, let us prove (iii). We know that for small $r$, the excited
states of $H_C$ tend to cluster in pairs. The eigenvalues from the odd
parity sector are independent of $r$, while those from the even parity
sector will converge from above to the odd ones (see Proposition
\ref{nedege1}). Consider such a pair of eigenvalues, which will always
remain separated from the rest of the spectrum if $r<r_1$ and $r_1$ is small
enough. Then we can find a contour $\Gamma$ which contains them and
$\inf_{z\in \Gamma}d_C(z)$ is bounded from below uniformly in
$r<r_1$. Then we can again write an estimate like \eqref{vetildar216},
and then apply the regular perturbation theory. The proof is over.

\end{proof}

\section{Reduction of $\widetilde{H^r}$ to
  $\Heff^r$}\label{section4}

We are now ready to go back to \eqref{newhs}, and argue why only the 
diagonal entries of the infinite operator-valued matrix 
$\{H_{m,n}\}_{m,n\in\Z}$ are important for the low lying spectrum of
$\widetilde{H^r}$. 

Let us formally write $\widetilde{H^r}$ as:
$$\widetilde{H^r}=H_{\rm diag}+V_{\rm offdiag},$$
where $H_{\rm diag}=\bigoplus_{n\in \Z}(H_{\eff}^r +\frac{n^2}{2r^2})$,
and $V_{\rm offdiag}$ contains all the non-diagonal entries of the
form $V_{m,n}^r$, $m\neq n$, (see \eqref{vemn}), and zero on the
diagonal. We will prove in this section that $V_{\rm offdiag}$ 
is relatively form bounded with respect to $H_{\rm diag}$, and
moreover, it is a "small" perturbation when $r$ is small. 

The main result is very similar to Proposition \ref{propohasc}, where
we only have to change $H_C$ by $H_{\rm diag}$ and $H_{\eff}^r$ by 
$\widetilde{H^r}$. Parity here only refers to the $x$ variable.  
Therefore we will start comparing the two operators. Before that, let
us note that the negative spectrum of $H_{\rm diag}$ is given by 
$H_{\eff}^r$ if $r$ is small enough. 

\subsection{$V_{\rm offdiag}$ is $H_{\rm diag}$-form bounded}

Let $\lambda_r=K\ln^2(r)$ with $K$ large enough and $r<r_0$, as in the
previous section. We know that $-\lambda_r\in\rho(H_{\rm diag})$, and
denote by $R_{\rm diag}(-\lambda_r)$ the resolvent $(H_{\rm
  diag}+\lambda_r)^{-1}$. Then the main technical result of this
subsection will be the following estimate: there exists $\delta>0$ and
$r_0(\delta)$ 
such that
\begin{equation}\label{margina55}
\left \Vert R_{\rm diag}^{1/2}(-\lambda_r)V_{\rm offdiag}
R_{\rm diag}^{1/2}(-\lambda_r)\right \Vert_{B_\infty(l^2(\Z;L^2(\R)))}=
\mathcal{O}(r^\delta\; \lambda_r^{-1/2}),\quad r<r_0.
\end{equation}
Let us first notice that we can replace $R_{\rm diag}^{1/2}$ by a simpler
operator, namely $\bigoplus_{n\in \Z}(\epsilon p_x^2 +1+ \frac{n^2}{2
  r^2})^{-1/2}$, where $\epsilon$ is a small enough positive
number. Indeed, we can write 
$$\langle f, [H_{\eff}^r+\lambda_r+  n^2/(2 r^2)]f\rangle \geq 
\langle f,[\epsilon p_x^2+\lambda_r/2+ n^2/(2 r^2) ]f\rangle,$$
where we used that for $\epsilon$ small enough we can show that: 
$$(1/2-\epsilon) p_x^2 -V_{\eff}^r+\lambda_r/2\geq 0,\quad r<r_0.$$
This means that 
\begin{equation}\label{margina56}
\left \Vert 
[\epsilon p_x^2+\lambda_r/2+ n^2/(2 r^2)]^{1/2}[H_{\eff}^r+
\lambda_r+  n^2/(2 r^2)]^{-1/2}\right \Vert_{B_\infty(L^2(\R))}\leq 1. 
\end{equation}
Define the bounded operators in $L^2(\R)$ (see \eqref{vemn}): 
\begin{align}\label{margina57}
\tilde{V}_{m,n}^r&:=
[\epsilon p_x^2+\lambda_r/2+ m^2/(2 r^2)]^{-1/2}V_{m,n}^r 
[\epsilon p_x^2+\lambda_r/2+ n^2/(2 r^2)]^{-1/2}, m\neq n,\nonumber \\
\tilde{V}_{m,m}^r&:=0,\quad m\in\Z. 
\end{align}
Then \eqref{margina55} would be implied by the following, stronger
estimate:
\begin{equation}\label{margina58}
\left \Vert \left \{\tilde{V}_{m,n}^r\right \}_{m,n\in\Z}
\right \Vert_{B_\infty(l^2(\Z;L^2(\R)))}=
\mathcal{O}(r^{\delta} \lambda_r^{-1/2}),\quad r<r_0.
\end{equation}
By an easy application of the Schur-Holmgren lemma, one can prove the estimate:
\begin{equation}\label{margina59}
\left \Vert \left \{\tilde{V}_{m,n}^r\right \}_{m,n\in\Z}
\right \Vert_{B_\infty(l^2(\Z;L^2(\R)))}\leq \sup_{m\in\Z}\sum_{n\in
  \Z}||\tilde{V}_{m,n}^r||_{B_\infty(L^2(\R))}.
\end{equation}

We now concentrate on the norms
$||\tilde{V}_{m,n}^r||_{B_\infty(L^2(\R))}$ and study their behavior
in $r$, $m$, and $n$. Remember that only the case $m\neq n$ is of
interest, since the diagonal terms are zero. 

Before anything else, let us do a unitary rescaling of $L^2(\R)$ by 
$(Uf)(x):=r^{1/2}f(rx)$ and $(U^*f)(x):=r^{-1/2}f(x/r)$. Then due to
various homogeneity properties we get:
 \begin{equation}\label{margina60}
U\tilde{V}_{m,n}^rU^*=r\cdot [\epsilon p_x^2+r^2\lambda_r/2+ m^2/2 ]^{-1/2}V_{m,n}^1
[\epsilon p_x^2+r^2\lambda_r/2+ n^2/2 ]^{-1/2}.
\end{equation}

We first give an important estimate for $V_{m,n}^1$, stated in the
next lemma:
\begin{lemma}\label{margina61}
Let $0<\alpha<1$ and $|m-n|\geq 1$. Fix any $0<\epsilon <1$. Then there exists a constant
$C=C(\alpha,\epsilon)$ such that we have the following
estimate:
\begin{equation}\label{margina62}
|V_{m,n}^1|(x)\leq C \left \{\frac{1}{|n-m|^\alpha}\;\frac{1}{
 |x|^{\alpha}}+\frac{1}{|n-m|}\right \},\quad |x|\leq r^{-\epsilon},
\end{equation}
and 
\begin{equation}\label{margina622}
|V_{m,n}^1|(x)\leq  {\rm const}\frac{r^{3\epsilon}}{|n-m|},\quad |x|\geq r^{-\epsilon}.
\end{equation}
\end{lemma}
\begin{proof} Due to symmetry properties we can write 
\begin{equation}\label{margina63}
V_{m,n}^1(x)=\frac{1}{2\pi}\int_{-\pi}^\pi \frac{\cos[(n-m)y]}
{[x^2+4\sin^2(y/2)]^{1/2}}dy.
\end{equation}
Integrating by parts we get:
\begin{equation}\label{margina64}
V_{m,n}^1(x)=\frac{1}{2\pi(n-m)}\int_{-\pi}^\pi \frac{\sin[(n-m)y]\sin(y)}
{[x^2+4\sin^2(y/2)]^{3/2}}dy.
\end{equation}
This equality immediately proves \eqref{margina622}. So we now focus
on $|x|\leq r^{-\epsilon}$. We can split the integral in two: one in which $|y|\geq \pi/2$,
and where the integrand has no singularities when $x$ is small, and
the second where $|y|\leq \pi/2$. In that region we can use the same
idea as in  Lemma \ref{logaritmicin} of replacing $\sin^2(y/2)$ by
$y^2$. We hence get:

\begin{equation}\label{margina65}
|V_{m,n}^1|(x)\leq \frac{{\rm const}}{|n-m|}\left (1+
\int_{-\pi/2}^{\pi/2} \frac{|\sin[(n-m)y]\sin(y)|}
{[x^2+y^2]^{3/2}}dy\right ).
\end{equation}
Now we employ the inequalities (here $0<\alpha<1$ is arbitrary):
$$|\sin[(n-m)y]|\leq |n-m|^{1-\alpha}|y|^{1-\alpha},\quad
|\sin(y)|\leq |y|,$$
then we make the change of variables $s=y/|x|$ and write:
\begin{equation}\label{margina66}
|V_{m,n}^1|(x)\leq \frac{{\rm const}}{|n-m|}\left (1+2
\frac{|n-m|^{1-\alpha}}{|x|^{\alpha}}
\int_{0}^{\infty} \frac{s^{2-\alpha}}
{[1+s^2]^{3/2}}ds\right ).
\end{equation}
Thus the lemma is proved.

\end{proof}

\vspace{0.5cm}

Now let us go back to \eqref{margina60}, and estimate the various
norms. If we write $V_{m,n}^1=V_{m,n}^1\chi(|\cdot|\leq
r^{-\epsilon})+V_{m,n}^1\chi(|\cdot|> r^{-\epsilon})$, 
then we have two different types of
estimates. When we keep $V_{m,n}^1\chi(|\cdot|> r^{-\epsilon})$, which is bounded,
then for the two resolvents we can use the usual $B_\infty(L^2)$ norm,
which together with \eqref{margina622} gives a contribution:
 
\begin{equation}\label{margina67}
\frac{{\rm const}\; r^{3\epsilon}}{\sqrt{r^2\lambda_r+n^2}\;\sqrt{r^2\lambda_r +m^2}\;|m-n|},\quad n\neq m.
\end{equation}

When we keep $V_{m,n}^1\chi(|\cdot|\leq  r^{-\epsilon})$, the estimate from 
\eqref{margina62} gives us that $[|V_{m,n}^1|\chi(|\cdot|\leq
1)]^{1/2}$ is an $L^2$ function, hence the operator 
$$\sqrt{|V_{m,n}^1|}\;\chi(|\cdot|\leq  r^{-\epsilon})[\epsilon p_x^2+r^2\lambda_r/2+ n^2/2 ]^{-1/2}$$
is Hilbert-Schmidt. Thus we have a product of two Hilbert-Schmidt
operators, and we can give an upper bound for the $B_\infty$
norm of their product of the form:  

\begin{equation}\label{margina68}
\frac{{\rm const(\alpha) }\;  r^{-\epsilon }}{(r^2\lambda_r+n^2)^{1/4}\;
(r^2\lambda_r +m^2)^{1/4}\;|m-n|^\alpha},\quad n\neq m.
\end{equation}

Therefore we obtained an upper bound for the norm of the operator in
\eqref{margina60} of the form:
\begin{align}\label{margina69}
||\tilde{V}_{m,n}^r||& \leq \frac{r\cdot {\rm
    const}\; r^{3\epsilon}}{\sqrt{r^2\lambda_r+n^2}\;\sqrt{r^2\lambda_r
    +m^2}\;|m-n|}\nonumber \\
&+\frac{r\cdot {\rm const }\cdot r^{-\epsilon}}{(r^2\lambda_r+n^2)^{1/4}\;
(r^2\lambda_r +m^2)^{1/4}\;|m-n|^\alpha},\quad m\neq n.
\end{align}

Remember that one is interested in the right hand side of
\eqref{margina59}. Now choose $1/2<\alpha<1$. 
We have to investigate several cases:

\begin{enumerate}
\item When $m=0$ and $|n|\geq 1$. Then 
  the first term in \eqref{margina69} will behave like
  $\lambda_r^{-1/2}\ r^{3\epsilon}|n|^{-2}$. 

The second term will behave like
  $r^{1/2-\epsilon}\lambda_r^{-1/4}|n|^{-1/2-\alpha}$. Both contributions are
  summable with respect to $n$. Note that if $\epsilon$ is small
  enough, both exponents of $r$ are positive. Denote by $\delta$ the
  smaller one. 

\item Fix $m\neq 0$, and consider all $n\neq m$. When $n=0$, we get
  similar terms as above. If $n\neq 0$,
  then we remain with the problem of summing up something like 
$$\sup_{m\neq 0}|m|^{-1/2}
\sum_{n\neq 0,n\neq
  m}\frac{1}{|n|^{1/2}|n-m|^{\alpha}},\;1/2<\alpha.$$ 
We can either use H\"older's inequality, or we can 
split the above sum in the following way: 
\begin{align}
&\sum_{n\neq 0,n\neq
  m}\frac{1}{|n|^{1/2}|n-m|^{\alpha}}\nonumber \\
& =
\left (\sum_{n\neq 0,n\neq
  m,|n|\leq |n-m| }+
\sum_{n\neq 0,n\neq
  m,|n|> |n-m| }\right )\frac{1}{|n|^{1/2}|n-m|^{\alpha}}\nonumber \\
&\leq \sum_{n\neq 0,n\neq
  m}\left
  (\frac{1}{|n|^{\alpha+1/2}}+\frac{1}{|n-m|^{1/2+\alpha}}\right )\leq
  {\rm const}(\alpha).
\end{align}

\end{enumerate}

We therefore consider \eqref{margina55} as proved. 

\subsection{Comparison between $\widetilde{H^r}$ and $H_{\rm
  diag}$}

If $r$ is small enough, we have the identity:
$$(\widetilde{H^r}+\lambda_r)^{-1}=R_{\rm diag}^{1/2}(-\lambda_r)
\left \{{\rm Id}+R_{\rm diag}^{1/2}(-\lambda_r)V_{\rm offdiag}R_{\rm
    diag}^{1/2}(-\lambda_r)\right \}^{-1}
R_{\rm diag}^{1/2}(-\lambda_r).$$
Moreover, this implies:
\begin{align}
||(\widetilde{H^r}+\lambda_r)^{-1}-R_{\rm diag}(-\lambda_r)||&\leq
\frac{{\rm const}}{\lambda_r} ||R_{\rm diag}^{1/2}(-\lambda_r)V_{\rm offdiag}R_{\rm
    diag}^{1/2}(-\lambda_r)||\nonumber \\
&\leq {\rm const}\frac{r^\delta\lambda_r^{-1/2}}{\lambda_r}.
\end{align}
This is the same as what we had in \eqref{vetildar209}, but with
$\lambda_r^{-1/2}\; r^\delta$ instead of $r^{4/9}$. Therefore we
can repeat the arguments of Proposition \ref{propohasc} and prove a
similar kind of spectrum stability for $\widetilde{H^r}$ and $H_{\rm
  diag}$.

\section{The main theorem and some conclusions}

We now try state a concentrated main result of our
paper. Let us first go back to the very first Hamiltonian which was
declared to be relevant for the exciton problem. This is 
$\bar{H}^r$, written in \eqref{pricipessa}. Because of the Coulomb
singularity, the best way to look at the spectral problem is to
consider its form $t_H$, given by \eqref{cinetica}, \eqref{teve} and
\eqref{formah}. We then managed to 
separate the mass center motion in the longitudinal direction, and we
got a simpler form $t_h$ in \eqref{primaschi2}. The center of the mass
cannot be separated in the transverse direction because of the
cylindrical geometry, but at least we can write $t_h$ as a direct sum of
$\bigoplus_{k\in\Z}t_{h_k}$. A crucial observation has been stated in
\eqref{inegali5}, which says that only $t_{h_0}$ is responsible for
the lowest lying spectrum of the original form. 

This gave us the possibility of renaming $t_{h_0}$ with
$\widetilde{t_H}$ in \eqref{tetilda}, and declare it as the central
object of study. Then in Proposition \ref{VR0compact} we constructed
its associated self-adjoint operator $\widetilde{H^r}$, where we had
to take care of the Coulomb-type singularity in two dimensions. 

Then after a unitary transformation induced by the discrete Fourier
transform with respect to the $y$ variable, we can see
$\widetilde{H^r}$ as an infinite operator valued matrix acting on the
Hilbert space $l^2(\Z; L^2(\R))$. We then decomposed $\widetilde{H^r}$
as the sum of a diagonal operator $H_{\rm diag}$ and an off-diagonal
part $V_{\rm offdiag}$. Eventually we proved in Section \ref{section4}
that the low lying spectrum of $\widetilde{H^r}$ is only slightly
influenced by the off-diagonal part for small $r$, and therefore the
relevant object remains $H_{\rm diag}$. 

But this diagonal part has the nice feature that each of its entry is of the
form $H_{\eff}^r+\frac{n^2}{2r^2}$, $n\in\Z$, where $H_{\eff}^r$ is
given in \eqref{hasefectiv} and \eqref{potefectiv}. Then in Section
\ref{section3}, more precisely in Proposition \ref{propohasc} 
we prove that the low lying spectrum of $H_{\eff}^r$ is
well approximated by the spectrum of a solvable operator, $H_C$, which
we discussed in Proposition \ref{nedege1}. 

We are now ready to collect all these results in the main theorem of our paper: 

\begin{theorem}\label{mainthm} 
The following three statements hold true:

{\rm (i)}. Fix $a<0$, and denote by $A:=\sigma(H_C)\cap
  (-\infty,a]$ and $B:=\sigma(\widetilde{H^r})\cap
  (-\infty,a]$. With the definition introduced in \eqref{esepsilon},
  we have that for every $\epsilon>0$, there exists
  $r_\epsilon>0$ such that for every $r<r_\epsilon$ we have
\begin{equation}\label{dist333f}
A\subset B_\epsilon,\quad B\subset A_\epsilon.
\end{equation}

{\rm (ii)}. The ground-state of $\widetilde{H^r}$ is non-degenerate, and diverges to 
$-\infty$ when $r\to 0$. The corresponding eigenfunction 
has even parity with respect to both variables. Moreover:

\begin{equation}\label{dist334f}
\lim_{r\to 0}|\inf\sigma(H_C)-\inf\sigma(\widetilde{H^r})|=0.
\end{equation}

{\rm (iii)}. Fix a compact interval $[a,b]\subset (-\infty,0)$ and
suppose that $H_C$ has exactly one eigenvalue $E_C$
in $[a,b]$, of parity $p=\pm$, for all $r<r_0$. 
Then if $r$ is small enough, $\widetilde{H^r}$ has exactly one
eigenvalue $\tilde{E}$ in this interval and 
$$\lim_{r\to 0}|\tilde{E}-E_C|=0.$$
Moreover, the corresponding eigenfunction has parity $p$ with respect to $x$.
\end{theorem}

Another important aspect of this problem is to determine
how fast these limits are assumed. We have not touched this issue
here, but we will study the numerical and physical implications of our
results in a consequent paper.   

\vspace{0.5cm}

\noindent {\bf Acknowledgments.} The authors thank T.G. Pedersen for
many fruitful discussions. H.C. was partially supported by the embedding
grant from {\it The Danish National Research Foundation: Network in
Mathematical Physics and Stochastics}. H.C. acknowledges support from the Danish 
F.N.U. grant {\it  Mathematical Physics and Partial Differential
  Equations}, and partial support through the European Union's IHP
network Analysis $\&$ Quantum HPRN-CT-2002-00277.

\end{document}